# Quality or Quantity: Toward a Unified Approach for Multi-organ Segmentation in Body CT

Fakrul Islam Tushar[*,a,b], Husam Nujaim[*,a,c], Wanyi Fu[a,b], Ehsan Abadi[a,b], Maciej A. Mazurowski[a,b], Ehsan Samei[a,b], William P. Segars[a,b], Joseph Y. Lo[a,b]

* These co-first authors contributed equally to this work.

[a] Center for Virtual Imaging Trials, Carl E. Ravin Advanced Imaging Laboratories, Dept. of Radiology, Duke University School of Medicine

[b] Department of Electrical and Computer Engineering, Pratt School of Engineering, Duke University, Durham, NC

[c] Erasmus+ Joint Master in Medical Imaging and Applications, Universitat de Girona, Spain; Université de Bourgogne, France; Università degli studi di Cassino, Italy.

## ABSTRACT

Organ segmentation of medical images is a key step in virtual imaging trials. However, organ segmentation datasets are limited in terms of *quality* (because labels cover only a few organs) and *quantity* (since case numbers are limited). In this study, we explored the tradeoffs between quality and quantity. Our goal is to create a unified approach for multi-organ segmentation of body CT, which will facilitate the creation of large numbers of accurate virtual phantoms.

Initially, we compared two segmentation architectures, 3D-Unet and DenseVNet, which were trained using XCAT data that is fully labeled with 22 organs, and chose the 3D-Unet as the better performing model. We used the XCAT-trained model to generate pseudo-labels for the CT-ORG dataset that has only 7 organs segmented. We performed two experiments: First, we trained 3D-UNet model on the XCAT dataset, representing *quality* data, and tested it on both XCAT and CT-ORG datasets. Second, we trained 3D-UNet after including the CT-ORG dataset into the training set to have more *quantity*. Performance improved for segmentation in the organs where we have true labels in both datasets and degraded when relying on pseudo-labels. When organs were labeled in both datasets, Exp-2 improved Average DSC in XCAT and CT-ORG by 1. This demonstrates that quality data is the key to improving the model's performance.

## INTRODUCTION

Segmentation of organs is an important requirement for many analyses related to virtual imaging trials including detection [1], classification [2] , or generating computational human phantoms [3, 4]. Due to the groundbreaking performance of the deep learning models, several approaches for multi-organ segmentation have been proposed [5, 6]. However, many of these approaches involved either a limited number of organs or more than one model trained for different organs. In a recent study [3], we introduced a 3D-Unet for organ segmentation using only 50 cases from the 4D extended cardiac-torso (XCAT) phantom series [7]. We segmented a comprehensive list of 22 organs from chest abdomen-pelvis (CAP) computed tomography (CT) scans [3]. In another study presented at SPIE previously [8], we used multiple public datasets to train separate models and ensemble the predictions together, but the excessive complexity of that approach made it very difficult to ensure the model performance was optimized and generalizable.

Some deep-learning practitioners believe that quantity or more training data improves the performance, such as by using pseudo-labels or weak supervision. However, others assert that quality data improves the model better. In this study, we investigated this critical issue through multi-organ segmentation. This work has two related aims. First, we seek to develop a unified framework for multi-organ 3D segmentation of CT images, which can be applied to and evaluated upon both public and private datasets. Second, we examine the effects of quality labels versus pseudo-labeling, which will guide the optimal use of these mixed datasets to improve performance to be more robust and accurate.

## METHODS

### Datasets:

We used the XCAT [7] phantoms dataset and CT-ORG [9] public dataset. The XCAT dataset consists of 50 adult CAP CT scans with 22 different organs labels: thyroid, lungs (L/R), heart, liver, spleen, kidneys (L/R), gallbladder, ribs (L/R), bladder, spine, clavicles, sternum, scapulas, stomach, pancreas, pelvis, femurs, arms, and body. 5-fold cross-validation is performed using random 60%/20%/20% split of train/validation/test set respectively. The same folds were used for all the validation. CT-ORG consists of 140 CT cases with 7 organ labels: lungs, bones, liver, kidneys, and bladder. 20 CT cases were kept as holdout test cases and excluded 9 PET-CT cases that were very different in resolution.

### Study Design:

As shown in figure 1, we first trained a 3D-UNet[3] and 3D DenseVNet [5] on the XCAT dataset. Based on the comparative performance, we decided to proceed with further experiments and analysis using 3D-UNet. We used that model to segment the CT-ORG dataset and thus generate pseudo labels for the 15 missing organs. Lastly, we trained a 3D-UNet model under the same framework on both datasets, XCAT and CT-ORG. We then compared the segmentation results of the model trained on the quality data (XCAT) and the model trained on more data (XCAT plus CT-ORG).

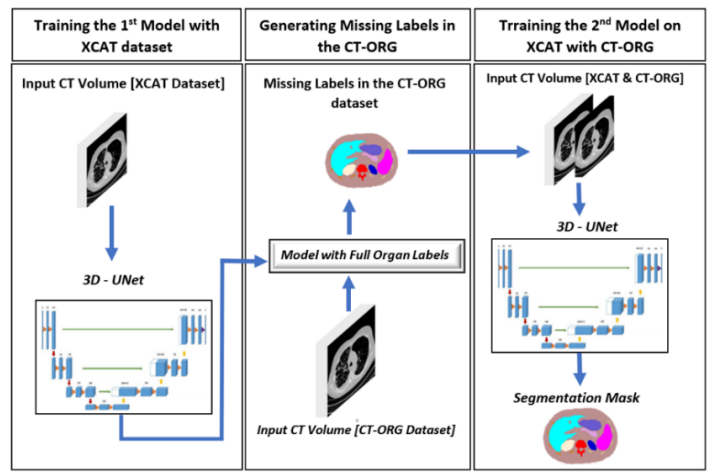

**Figure 1.** Study design flowchart.

## Experiments:

First, we trained a 3D-UNet and 3D DenseVNet with the recommended hyper-parameters suggested by the authors in the literature [3, 5] for 5-folds cross-validation. Patch-wise training and a sliding window-based inference scheme have been used for the experiments. Before training the model, each volume was resampled to 2.5×2.5×5 mm (x, y, z), and the Hounsfield unit window used is between -1000 to 800. The intensity values were normalized to [-1, 1]. 32 patches each at 128x128x128 were extracted per volume for all the experiments. Training took approximately took 48 hours using a 24 GB Nvidia TITAN RTX GPU. Inference time was <20 sec per volume.

Further, we have conducted 2 experiments, as detailed in Table 1. In Experiment-1, we trained on the XCAT dataset and validated it on the same dataset. In Experiment-2, we trained on both XCAT plus CT-ORG datasets. Both experiments were validated on the same XCAT subset and internally tested on the holdout XCAT and CT-ORG subset. Due to the computational time, only fold-1 of 5 folds is reported, but we will show full cross-validation results at the conference.

**Table 1**. Details of the conducted experiments.

| Experiments | Training Dataset | Validation Dataset | Test Dataset |
|---|---|---|---|
| Experiment-1 | XCAT (n=30) | XCAT (n=10) | XCAT (n=10), CT-ORG (n=20) |
| Experiment-2 | XCAT (n=30), CT-ORG (n=111) | | |

## RESULTS and DISCUSSION

The Dice similarity coefficient (DSC) metric for each organ was calculated to assess the accuracy of both segmentation models. Our experiments show that the UNet model is significantly superior to the DenseVNet model in the multi-organ segmentation task, as illustrated in Figure 2. The input data is exactly the same for both models, including same cases and same organ labels. Although both models may be acceptable, the UNet model is faster and easier to converge.

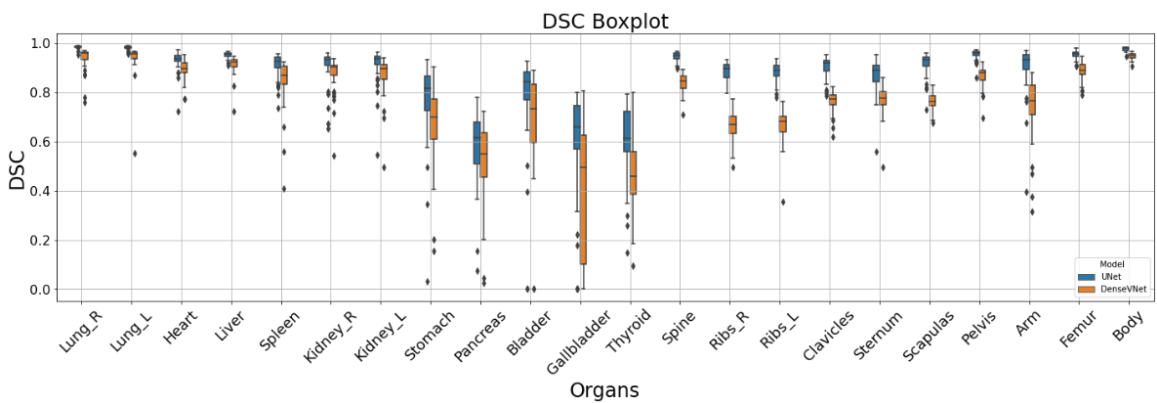

**Figure 2.** Illustrates the DSC comparison between UNet (blue) and DenseVNet (orange).

For organs labeled in both datasets, Experiment-2 was the same or slightly better than Experiment-1 in the XCAT dataset, such as the bladder improving from 0.807 to 0.870. On the other hand, the use of the pseudo labels in Experiment-2 degraded the performance for XCAT organs that do not have labels in the CT-ORG dataset, such as thyroid dropping from 0.550 to 0.465. Figure 3 shows an example of the segmentation of the two experiments from the XCAT test set. Table 2 shows the average DSC of the two experiments.

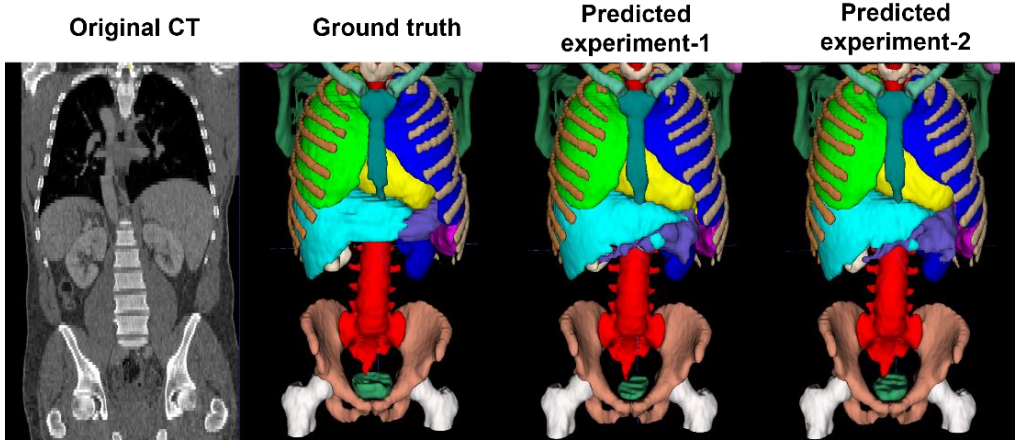

**Figure 3.** Example of segmentation results. In this case, the bladder (dark green at the bottom) improved from DSC=0.500 for Experiment-1 to 0.660 for Experiment-2

**Table 2.** Average Dice similarity coefficient for the 2 experiments. Missing labels in CT-ORG are shown as "---" symbol.

| Organ | Experiment one | | Experiment two | |
|---|---|---|---|---|
| | XCAT | CT-ORG | XCAT | CT-ORG |
| Lung-L | 0.982 | 1.000 | 0.985 | 0.987 |
| Lung-R | 0.982 | 1.000 | 0.983 | 0.98 |
| Heart | 0.898 | --- | 0.900 | --- |
| Liver | 0.954 | 0.918 | 0.956 | 0.949 |
| Spleen | 0.896 | --- | 0.908 | --- |
| Kidney-R | 0.887 | 0.833 | 0.902 | 0.917 |
| Kidney-L | 0.858 | 0.822 | 0.862 | 0.908 |
| Stomach | 0.656 | --- | 0.654 | --- |
| Pancreas | 0.516 | --- | 0.505 | --- |
| Bladder | 0.807 | 0.727 | 0.870 | 0.858 |
| Gallbladder | 0.592 | --- | 0.536 | --- |
| Thyroid | 0.550 | --- | 0.465 | --- |
| Spine | 0.942 | 1.000 | 0.939 | 0.938 |
| Ribs_R | 0.890 | 0.995 | 0.897 | 0.923 |
| Ribs_L | 0.892 | 1.000 | 0.894 | 0.921 |
| Clavicles | 0.890 | --- | 0.829 | --- |
| Sternum | 0.836 | --- | 0.803 | --- |
| Scapulas | 0.895 | --- | 0.886 | --- |
| Pelvis | 0.953 | --- | 0.950 | --- |
| Arm | 0.871 | --- | 0.811 | --- |
| Femur | 0.946 | --- | 0.946 | --- |
| Body | 0.973 | --- | 0.974 | --- |

## LIMITATION AND FUTURE WORK

Both models (UNet & DenseVNet) showed weaker performance in the segmentation of a few traditionally difficult organs such as the pancreas, thyroid, gallbladder, and stomach. We will improve the segmentation of these organs by focusing on datasets where they are annotated and by using semiautomated labeling. Our future work includes quantifying the quality of generated pseudo-labels utilizing the uncertainty quantification and analyzing different effects of imaging properties on model predictions utilizing virtual imaging trials including.

## NEW & BREAKTHROUGH WORK

1) We directly compared two segmentation architectures used in prior work, namely 3D-UNet [4] and 3D DenseVNet [5] models on the XCAT [7] dataset. The better-performing 3D-UNet was picked for further analysis.
2) We trained the same architecture with the CT-ORG [9] dataset that was pseudo-labeled in a semi-supervised manner and the XCAT datasets and compared their segmentation comparison directly.

## ACKNOWLEDGMENTS

This work was funded in part by the Center for Virtual Imaging Trials, NIH/NIBIB P41-EB028744. Husam Nujaim performed part of this work as a visiting student from Medical Imaging and Applications (MAIA) and holds an EACEA Erasmus+ grant.

## REFERENCES

[1] P. Schelb *et al.*, "Classification of Cancer at Prostate MRI: Deep Learning versus Clinical PI-RADS Assessment," *Radiology,* vol. 293, no. 3, pp. 607-617, Dec 2019, doi: 10.1148/radiol.2019190938.
[2] F. I. Tushar *et al.*, "Classification of Multiple Diseases on Body CT Scans Using Weakly Supervised Deep Learning," *Radiology: Artificial Intelligence,* vol. 4, no. 1, p. e210026, 2022, doi: 10.1148/ryai.210026.
[3] W. Fu *et al.*, "iPhantom: A Framework for Automated Creation of Individualized Computational Phantoms and Its Application to CT Organ Dosimetry," *IEEE J Biomed Health Inform,* vol. 25, no. 8, pp. 3061-3072, Aug 2021, doi: 10.1109/JBHI.2021.3063080.
[4] W. Fu *et al.*, "iPhantom: an automated framework in generating personalized computational phantoms for organ-based radiation dosimetry," presented at the Medical Imaging 2021: Physics of Medical Imaging, 2021.
[5] E. Gibson *et al.*, "Automatic multi-organ segmentation on abdominal CT with dense v-networks," *IEEE Trans Med Imaging,* vol. 37, no. 8, pp. 1822-1834, Aug 2018, doi: 10.1109/TMI.2018.2806309.
[6] O. Ronneberger, P. Fischer, and T. Brox, "U-Net: Convolutional Networks for Biomedical Image Segmentation," Cham, 2015: Springer International Publishing, in

Medical Image Computing and Computer-Assisted Intervention – MICCAI 2015, pp. 234-241.
[7] W. P. Segars *et al.*, "Population of anatomically variable 4D XCAT adult phantoms for imaging research and optimization," *Medical Physics,* vol. 40, no. 4, p. 043701, 2013, doi: 10.1118/1.4794178.
[8] W. Fu *et al.*, *Multi-organ segmentation in clinical-computed tomography for patient-specific image quality and dose metrology* (SPIE Medical Imaging). SPIE, 2019.
[9] K. S. Blaine Rister, Tomomi Nobashi and Daniel L. Rubin., "CT-ORG: CT volumes with multiple organ segmentations [Dataset]," 2019, doi: DOI: 10.7937/tcia.2019.tt7f4v7o.